\documentclass{aastex631}

\usepackage{amsmath}
\usepackage{verbatim}
\graphicspath{{./}{figures/}}
\begin{document}
\title{A Tale of Three: Magnetic Fields along the Orion Integral-Shaped Filament as Revealed by JCMT BISTRO survey}
\author[0000-0001-7276-3590]{Jintai Wu}
\affiliation{School of Astronomy and Space Science, Nanjing University, 163 Xianlin Avenue, Nanjing 210023, People's Republic of China}
\affiliation{Key Laboratory of Modern Astronomy and Astrophysics (Nanjing University), Ministry of Education, Nanjing 210023, People's Republic of China}

\author[0000-0002-5093-5088]{Keping Qiu}
\affiliation{School of Astronomy and Space Science, Nanjing University, 163 Xianlin Avenue, Nanjing 210023, People's Republic of China}
\affiliation{Key Laboratory of Modern Astronomy and Astrophysics (Nanjing University), Ministry of Education, Nanjing 210023, People's Republic of China}

\author[0000-0002-5391-5568]{Fr\'{e}d\'{e}rick Poidevin}
\affiliation{Instituto de Astrofisíca de Canarias, E-38205 La Laguna,Tenerife, Canary Islands, Spain}
\affiliation{Universidad de La Laguna, Dpto. Astrofísica, E-38206 La Laguna, Tenerife, Spain}

\author[0000-0002-0794-3859]{Pierre Bastien}
\affiliation{Centre de recherche en astrophysique du Qu\'{e}bec \& d\'{e}partement de physique, Universit\'{e} de Montr\'{e}al,1375, Avenue Thérèse-Lavoie-Roux, Montréal, QC, H2V OB3, Canada}

\author[0000-0002-4774-2998]{Junhao Liu}
\affiliation{Division of ALMA, National Astronomical Observatory of Japan, Mitaka, Tokyo 181-8588, Japan}

\author[0000-0001-8516-2532]{Tao-Chung Ching}
\affiliation{Zhejiang Lab, Kechuang Avenue, Yuhang District, Hangzhou 311121, People's Republic of China}

\author[0000-0001-7491-0048]{Tyler L. Bourke}
\affiliation{SKA Observatory, Jodrell Bank, Lower Withington, Macclesfield SK11 9FT, UK}
\affiliation{Jodrell Bank Centre for Astrophysics, School of Physics and Astronomy, University of Manchester, Oxford Road, Manchester, M13 9PL, UK}

\author[0000-0003-1140-2761]{Derek Ward-Thompson}
\affiliation{Jeremiah Horrocks Institute, University of Central Lancashire, Preston PR1 2HE, UK}

\author[0000-0002-8557-3582]{Kate Pattle}
\affiliation{Department of Physics and Astronomy, University College London, WC1E 6BT London, UK}

\author[0000-0002-6773-459X]{Doug Johnstone}
\affiliation{NRC Herzberg Astronomy and Astrophysics, 5071 West Saanich Road, Victoria, BC V9E 2E7, Canada}
\affiliation{Department of Physics and Astronomy, University of Victoria, Victoria, BC V8W 2Y2, Canada}

% ...

\begin{abstract}
As part of the BISTRO survey, we present JCMT 850 $\mu$m polarimetric observations towards the Orion Integral-Shaped Filament (ISF) that covers three portions known as OMC-1,  OMC-2, and OMC-3. The magnetic field threading the ISF seen in the JCMT POL-2 map appears as a tale of three: pinched for OMC-1, twisted for OMC-2, and nearly uniform for OMC-3. A multi-scale analysis shows that the magnetic field structure in OMC-3 is very consistent at all the scales, whereas the field structure in OMC-2 shows no correlation across different scales. In OMC-1, the field retains its mean orientation from large to small scales, but shows some deviations at small scales. Histograms of relative orientations between the magnetic field and filaments reveal a bimodal distribution for OMC-1, a relatively random distribution for OMC-2, and a distribution with a predominant peak at 90$^\circ$ for OMC-3. Furthermore, the magnetic fields in OMC-1 and OMC-3 both appear to be aligned perpendicular to the fibers, which are denser structures within the filament, but the field in OMC-2 is aligned along with the fibers. All these suggest that gravity, turbulence, and magnetic field are each playing a leading role in OMC-1, 2, and 3, respectively. While OMC-2 and 3 have almost the same gas mass, density, and non-thermal velocity dispersion, there are on average younger and fewer young stellar objects in OMC-3, providing evidence that a stronger magnetic field will induce slower and less efficient star formation in molecular clouds.
\end{abstract}
\keywords{Star formation(1569); Interstellar magnetic fields(845); Interstellar clouds(834); Polarimetry(1278)}

\section{Introduction}
During the star formation process, the dynamics and physical states of the molecular clouds are influenced by various physical mechanisms, especially self-gravity, turbulence and magnetic field ($B$-field) \citep{2007ARA&A..45..565M}. It has long been a subject of intense debate as to which force is playing a dominant role in regulating the cloud collapse and fragmentation \citep{2004RvMP...76..125M,2006ApJ...646.1043M,2012ARA&A..50...29C}. Regarding the $B$-field, either the ``strong-field models" which support a defining role played by the $B$-field \citep[e.g.,][]{2006ApJ...646.1043M}, or the ``weak-field models" that pay more attention to turbulence \citep[e.g.,][]{2004RvMP...76..125M}, cannot sufficiently explain all the observations towards star formation regions. The relative importance of turbulence and $B$-field as well as their interactions with self-gravity in star formation remain to be explored in more case studies \citep{2021Galax...9...41L}. More reasonable scenarios may need to consider essential roles of both processes, which have been explored in simulations \citep{2012ARA&A..50...29C,2019FrASS...6....5H}.

Dense molecular filaments are important sites for star formation, with molecular gas accumulating and then fragmenting into star-forming cores due to gravitational instability \citep{2014prpl.conf...27A, 2023ASPC..534..233P,2023ASPC..534..153H}. Observations have shown that $B$-fields appear to be perpendicular to high-density filaments, while they appear to be parallel to low-density elongated clouds or striations \citep[e.g.,][]{2016A&A...590A.110C}. Magnetic fields may also play a central role in shaping the fragmentation and physical states of filaments \citep[e.g.,][]{2019ApJ...878...10T,2021A&A...647A..78A}. More observations and dedicated studies are needed to revealing the relative importance of $B$-field compared to other processes and to decipher how $B$-field influence the gas dynamics during filament formation and fragmentation.

Situated on the head of the Orion A giant molecular cloud, the Integral-Shaped Filament (ISF) is a well-known nearby star-forming filament \citep{1999ApJ...510L..49J,2008hsf1.book..459B} containing several portions, of which the more extensively studied are OMC-1, OMC-2, and OMC-3. Several studies present $B$-field results of the whole ISF \citep[e.g.,][]{2004ApJ...604..717H, 2009ApJS..182..143M}, or its portions OMC-1 \citep[e.g.,][]{2017ApJ...842...66W, 2019ApJ...872..187C, 2022EPJWC.25700002A}, OMC-2/3 \citep{2010ApJ...716..893P, 2022A&A...659A..22Z, 2022MNRAS.514.3024L}, and OMC-4 \citep{2022MNRAS.514.3024L}. With active massive star formation, the $B$-field in OMC-1 has been detected with a large-scale hourglass morphology associated with two molecular clumps, namely Orion BN/KL and South \citep[e.g.,][]{1998ApJ...493..811S,2017ApJ...842...66W,2017ApJ...846..122P}. The $B$-field orientations in OMC-2 exhibit more variations compared to the other portions of the ISF \citep[e.g.,][]{2010ApJ...716..893P}. As for OMC-3, observations have revealed a more ordered $B$-field \citep[e.g.,][]{2001ApJ...562..400M}. Therefore, being the nearest filamentary molecular cloud \citep[393 pc,][]{2018A&A...619A.106G} forming both massive and intermediate- to low-mass stars, the OMC-1/2/3 region shows hints of varying $B$-field properties along the ISF, and a more comprehensive investigation is expected to provide new insights into the role of $B$-fields in filament dynamics and star formation. In this current work, as part of the $B$-fields In Star-forming Region Observations \citep[BISTRO,][]{2017ApJ...842...66W,2020pase.conf..117B}, we use the James Clerk Maxwell Telescope (JCMT) to make submm polarimetric observations of the ISF. The BISTRO team has previously observed the ISF \citep{2017ApJ...842...66W,2017ApJ...846..122P}. However, those observations were focused only on OMC-1. In this paper we have more than doubled the area studied, to also include OMC-2 and 3. The aim is to set those earlier observations in the context of their environment and to understand the bigger picture of the role of magnetic fields in Orion A.

\section{Observations}\label{sec:obs}
The observations of polarized dust emission (project ID: M17BL011, M20AL018) covering the OMC-1, 2, and 3 in the Orion ISF were performed using POL-2 \citep{2016SPIE.9914E..03F} together with SCUBA-2 \citep{2013MNRAS.430.2513H} on the JCMT. Some observations (project ID: M15BEC02) toward OMC-1 south were taken during the POL-2 commissioning stage. All the data were obtained using the POL-2 DAISY mode \citep{2016SPIE.9914E..03F}. 

The reduction of raw data involves three primary steps and uses two packages SMURF and KAPPA \citep{2013ascl.soft10007J,2014ascl.soft03022C} in the Starlink package \citep{2014ASPC..485..391C}. With an effective beam size of 14.1$\arcsec$ ($\sim$0.027 pc at 393 pc) at 850 $\mu$m \citep{2013MNRAS.430.2534D}, we produced a synthesized map of Stokes parameters using a pixel size of 4$\arcsec$. We perform the absolute flux calibration with the flux conversion factor (FCF) estimated by adopting different recommended FCF values \citep{2021AJ....162..191M} weighted with the observation time. In our work, FCFs were set to 695 ${\rm Jy\ beam^{-1}\ pW^{-1}}$ for OMC-1, and to 668 ${\rm Jy\ beam^{-1}\ pW^{-1}}$ for OMC-2, 3.

By using three models including background, source, and residual components, we smoothed the polarization maps to revise some obviously inaccurate measurements due to the uncertainty. With the calibrated Stokes parameters, the polarized intensities ($PI$), polarization degrees ($P$), and angles ($\theta_P$) at different positions can be calculated using the following equations:

\begin{equation}
PI = \sqrt{Q^2 + U^2},\ \ P=\frac{PI}{I},\ \ {\rm and}\ \theta_P=0.5\tan^{-1}(\frac{U}{Q}).
\end{equation}

Since both positive and negative $Q$ and $U$ values contribute to a positive $PI$ value, a modified asymptotic estimator \citep[MAS,][]{2014MNRAS.439.4048P} is employed to de-bias the results to avoid the overestimation of $PI$. 

Using these equations, the parameters and their uncertainties are determined to produce a catalogue of polarization half-vectors. The polarization vectors with $P/\sigma_P<3$ or $\sigma_P>5\%$ are removed in our analysis, where $\sigma_P$ is the uncertainty of polarization degree. All the selected polarization vectors, with their lengths proportional to the polarization degrees and plotted in an interval of 8$\arcsec$, are all shown in Figure \ref{Bmap}(a).

\section{Results} \label{sec:resul}
\subsection{Magnetic field morphology} \label{subsec:morph}
\begin{figure*}[htbp]
\centering
\includegraphics[width=1.0\columnwidth]{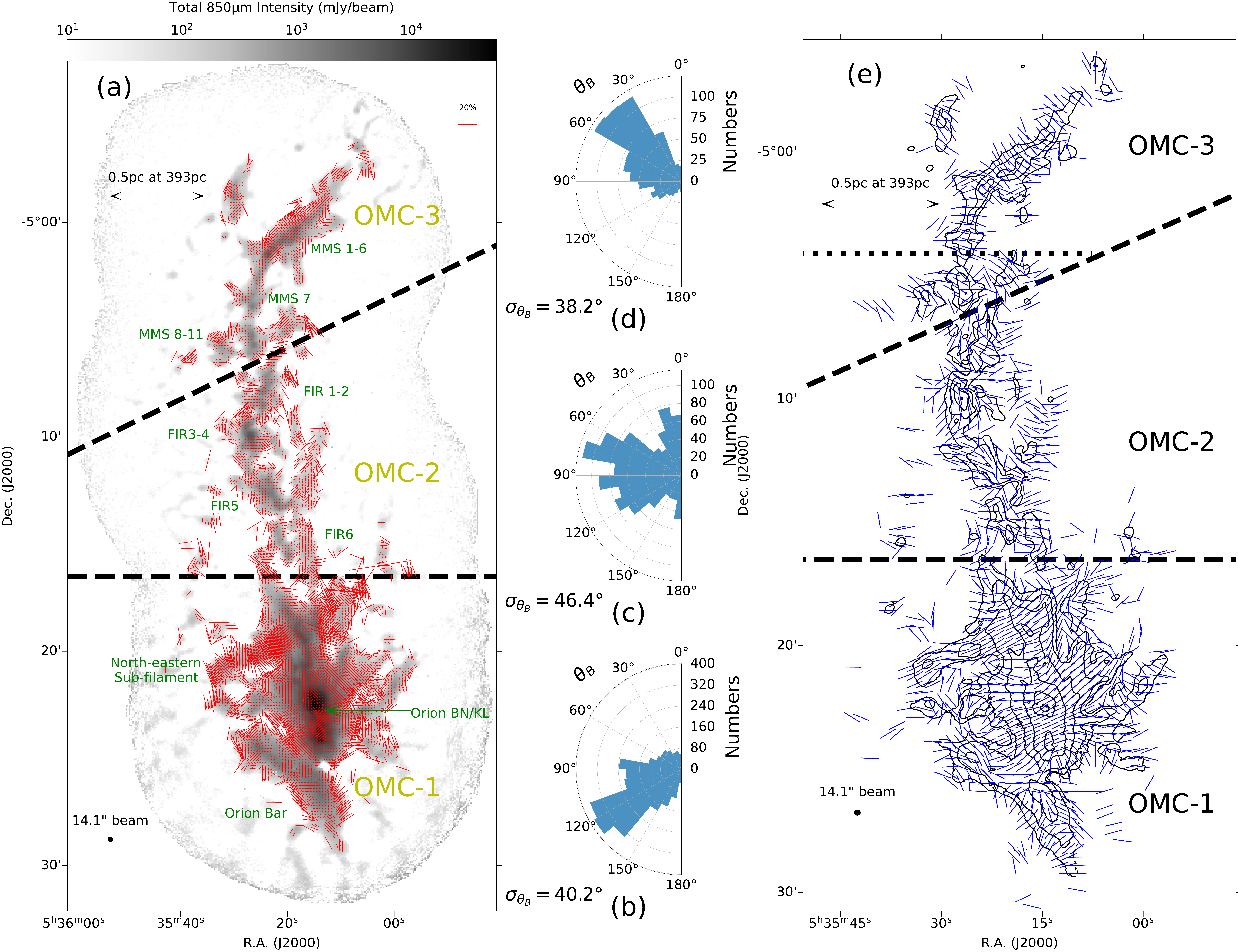}
\caption{Panel (a): Dust polarization observations of Orion A ISF made with the POL-2 on JCMT. Grayscale image shows the 850 $\mu$m total intensity (Stokes $I$). Red vectors are plotted in an interval of 8$\arcsec$, showing the polarization angles with the length proportional to the polarization degree. Black dashed lines mark the divisions between three clouds, i.e., OMC-1 to the south, OMC-2 in the middle, and OMC-3 to the north. The locations of the 1.3 mm sources identified by \citet{1997ApJ...474L.135C} including MMS 1-10 in OMC-3, FIR 1-6 in OMC-2, and the Northeastern Sub-filament and Orion Bar in OMC-1 are marked on the image. Panels (b), (c), and (d) show the histograms of the position angles of the $B$-field orientations for OMC-1, OMC-2, and OMC-3, respectively. Panel (e): Blue vectors with a uniform arbitrary length are plotted in an interval of 20$\arcsec$, showing the magnetic field orientations, and are derived by rotating the polarization vectors by $90^\circ$. The 850 $\mu$m total intensity is shown in black contours at $\rm log_{10}$ scale ($\rm mJy\ beam^{-1}$), which starts from 2.2 and continues at steps of 0.5}. The black dotted line splits the OMC-3 cloud into the North and South parts.
\label{Bmap}
\end{figure*}

Assuming aligned dust grains regulated by $B$-fields based on the Radiation Alignment Theory (RAT) \citep{2007JQSRT.106..225L}, the polarization angles of thermal dust emission enable one to infer the orientation of the $B$-field projected on the plane of sky. Figure \ref{Bmap}(b-d) shows the statistics of $B$-field orientation distributions of the three clouds. It is clear that the $B$-fields in OMC-1 are mostly aligned along a northwest-southeast orientation with a position angle (PA) of about 120$^\circ$, while the $B$-fields in OMC-3 are predominantly aligned along a northeast-southwest orientation with a PA of about 45$^\circ$. On the other hand, the $B$-field orientations in OMC-2 have a broad distribution between 50$^{\circ}$ and 130$^{\circ}$ and another group between 0$^{\circ}$ and 30$^{\circ}$, indicating a relatively more random distribution. In Figure \ref{Bmap}(e), we present the half-vectors rotated by 90$^\circ$ in an interval of 20$\arcsec$ representing the corresponding $B$-field orientations across the filament.

\textbf{OMC-1: }Overall, the $B$-field appears to be perpendicular to the main axis of the cloud/filament. OMC-1 is associated with the Orion Nebula Cluster (ONC) and contains a high concentration of gas at a high temperature of $>$100K \citep[e.g., ][]{2020ApJ...901...62L}. The maximum 850$\mu$m brightness of OMC-1 is approximately $9\times10^5$ mJy/beam, and is associated with the hot, high-mass star-forming clumps of Orion BN/KL. Moreover, as one approaches the location of Orion BN/KL, the hourglass pattern of the $B$-field becomes more prominent. This pinched morphology in the central cloud of OMC-1 indicates a strong interaction between gravity and $B$-field, showcasing the effects of the $B$-field in high-mass star forming regions. Figure \ref{Bmap} also shows $B$-field lines that are aligned parallel with the orientation of the cloud extension in the northeastern sub-filament of OMC-1, suggesting a gas accumulation process that is guided by the $B$-field surrounding the filament. 

\textbf{OMC-2: }Our dust polarization map of OMC-2 is a marked improvement over previous observations, such as the SCUPOL results \citep{2010ApJ...716..893P}. As a site for intermediate-to-low mass star formation, OMC-2 seems to have relatively chaotic $B$-field structures compared to the other two clouds. From Figure \ref{Bmap}(e), the $B$-field lines seem to converge towards denser areas in the central parts of OMC-2, where gravitational contraction is likely taking place. In contrast, a sub-filament to the west of the main filament is overall perpendicular to the $B$-field.

\textbf{OMC-3: }OMC-3 appears to be a filamentary cloud extending from southeast to northwest. The POL-2 observations reveal a nearly uniform $B$-field in the northern backbone of OMC-3. However, the OMC-3 South, which is suspected to be a "second filament" as noted by \citet{2010ApJ...716..893P}, has disorderly $B$-field directions that are similar to the complicated structures of OMC-2. In the main body of OMC-3 North, the $B$-field orientations are orthogonal to the filament direction. In brief, OMC-3 exhibits very ordered to even uniform $B$-field structures.

\subsection{Multi-scale view of B-field geometries in Orion ISF}
\begin{figure*}[htbp]
\centering
\includegraphics[width=0.9\columnwidth]{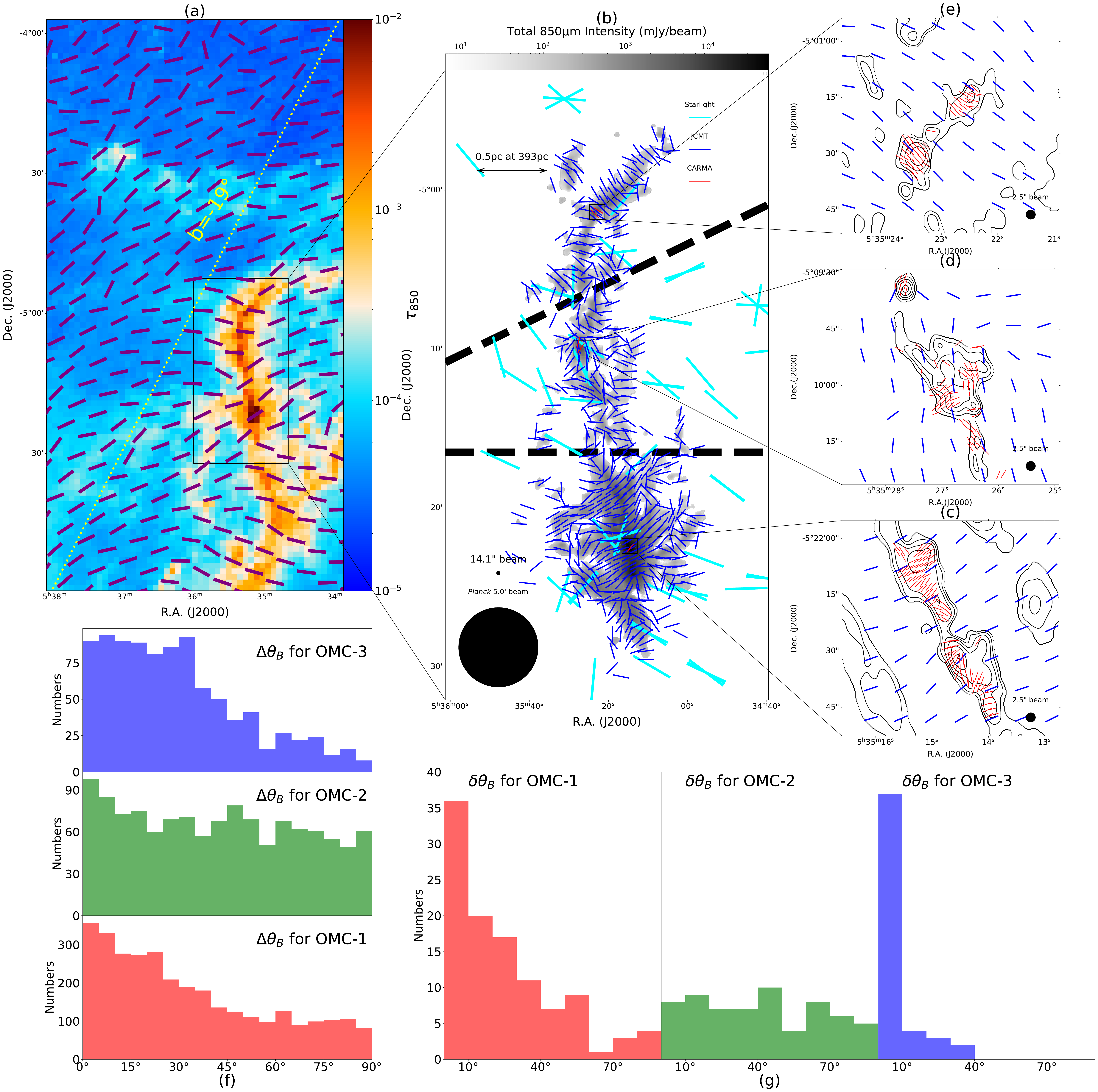}
\caption{Multi-scale $B$-field orientations in the ISF. Panel (a): the background image displays 850 $\mu$m opacity map obtained from the \emph{Herschel} and \emph{Planck} data \citep{2014A&A...566A..45L}; purple segments indicate the large-scale $B$-field orientations inferred from the \emph{Planck} 353 GHz data. A yellow dotted line marks the galactic latitude b=19$^\circ$. Panel (b): $B$-field orientations derived from starlight, JCMT POL-2, and CARMA observations. The background image shows the JCMT 850 $\mu$m total intensity map; blue vectors plotted at an interval of 32$\arcsec$ denote the $B$-field orientations observed by JCMT POL-2; cyan vectors represent the $B$-field orientations revealed by starlight polarization observations \citep{2011ApJ...741..112P}; red vectors show the averaged $B$-field orientations obtained by the CARMA TADPOL survey \citep{2014ApJS..213...13H}. In Panels (c), (d), and (e), red vectors indicate the $B$-field orientations obtained by the CARMA TADPOL survey in selected dense cores located in OMC-1, OMC-2, and OMC-3, respectively; blue vectors show the $B$-field orientations derived with the JCMT POL-2 observations; the CARMA observations of the total dust emission at 1.3 mm are shown in black contours at ${\rm log_{10}}$ scale (${\rm mJy\ beam^{-1}}$), which starts from -2.0 and continues at steps of 0.5 in Panel (c), starts from -2.0 and continues at steps of 0.2 in Panel (d), starts from -1.0 and continues at steps of 0.3 in Panel (e). Panel (f): Histograms of the difference angles between the $B$-field orientations unveiled by \emph{Planck} and that by the JCMT POL-2 for OMC-1, OMC-2, and OMC-3. Panel (g): Histograms of the difference angles between the $B$-field orientations obtained by the JCMT POL-2 and that by the CARMA TADPOL survey for OMC-1, OMC-2, and OMC-3.}
\label{figveccom}
\end{figure*}

We utilized the 353 GHz polarization observations made with the High Frequency Instrument (HFI) on \emph{Planck} to infer the large-scale $B$-field \citep{2015A&A...576A.104P}. The Stokes $I$, $Q$, and $U$ maps, which were corrected for the contamination from the Cosmic Microwave Background (CMB) and Cosmic Infrared Background (CIB), were used to generate the large-scale polarization map at a resolution of 5$\arcmin$. Figure \ref{figveccom}(a) displays the large-scale $B$-field maps around the Orion A region. The $B$-field is roughly perpendicular to the ISF. Moreover, the field structure appears slightly pinched toward the filament.

In addition, we check the optical starlight polarization observations to further explore the large-scale $B$-field in relatively low-density regions \citep{2011ApJ...741..112P}. To limit our analysis to sources within the Orion cloud, we only consider starlight detections that fall within the region of our JCMT observations and have a distance of 360$\sim$500 pc based on Gaia parallax measurements \citep{2020A&A...643A.151R, 2021A&A...649A...1G, 2021AJ....161..147B}. The $B$-fields derived from 61 detections are shown in Figure \ref{figveccom}(b). The majority of the $B$-field half-vectors have a west-east or northeast-southwest orientation, in general consistent with the \emph{Planck} results. Given the optical polarization data are presumably tracing the $B$-field threading the ISM around the ISF or that in the foreground toward the ISF, the $B$-field structure shows small deviation compared to that seen in the \emph{Planck} map.

The TADPOL survey \citep{2014ApJS..213...13H} mapped the $B$-fields toward several selected sources in the ISF, including Orion KL in OMC-1, FIR 3 and FIR 4 in OMC-2, and MMS 5 and MMS 6 in OMC-3, at an angular resolution of 2.5$\arcsec$ (0.005 pc) using the Combined Array for Research in Millimeter-wave Astronomy (CARMA). From Figure \ref{figveccom}(c-e), for OMC-1, the orientation of the small scale $B$-field revealed by the CARMA largely follows that of the intermediate scale $B$-field seen by the JCMT POL-2, though a small fraction of the CARMA $B$-field half-vectors are offset from parallel to even perpendicular to the JCMT $B$-field half-vectors; for OMC-2, the small scale $B$-field is apparently decoupled from that on the intermediate scale and there is no obvious correlation between the orientations of the $B$-fields on the two scales; for OMC-3, the small scale and intermediate scale $B$-fields both appear to be uniform with the almost same orientation. There have been new ALMA observations of dust polarization toward several sources in the ISF; however, these observations were made at sub-arcsec resolutions, either probing $B$-field structures at too small scales to be compared with the JCMT data \citep{2021ApJ...907...94C} or being dominated by self-scattering and thus unable to probe the $B$-field structure \citep{2019ApJ...872...70T,2024ApJ...963..104L}.

More quantitatively, we compare the intermediate scale $B$-field probed by JCMT POL-2 with the large scale $B$-field probed by \emph{Planck} by calculating the difference angle, $\Delta \theta_B$, between the orientations of the $B$-fields on the two scales. Since the \emph{Planck} map covers an area much larger than that is covered by the JCMT map, we calculate $\Delta \theta_B$ for each half-vector at $8\arcsec$ interval in the JCMT map; the $B$-field orientation at the corresponding position in the \emph{Planck} map is derived by a weighted average of the $B$-field orientations at the nearest four pixels, where the pixel size of the \emph{Planck} map is $2\arcmin$ and the weighting is taken as the inverse of the square of the distance between the pixel center to the position of interest. In Figure \ref{figveccom}(f), the histogram of $\Delta \theta_B$ for OMC-1 is clearly peaking toward 0$^\circ$, suggesting that the orientation of the intermediate scale (0.03 pc) $B$-field is predominantly parallel with that of the large scale (0.6 pc) $B$-field. For OMC-2, $\Delta \theta_B$ appears to be widely distributed between 0$^\circ$ and 90$^\circ$, with a very minor tendency of peaking at 0$^\circ$, indicating that the $B$-field orientation on the intermediate scale has shown strong local variation and started to decouple from that on the large scale. For OMC-3, $\Delta \theta_B$ are almost all below 35$^\circ$, indicating that the intermediate scale $B$-field is well aligned with that on the large scale.

Similarly to $\Delta \theta_B$, we compute the difference angle, $\delta \theta_B$, between the orientations of the $B$-fields probed by JCMT and CARMA, for each CARMA detection. Again the $B$-field orientation at the corresponding position in the JCMT map is derived by a weighted average of the $B$-field orientations at the nearest four pixels, and the weighting is taken as the inverse of the square of the distance between the pixel center to the position of interest. In Figure \ref{figveccom}(g): the distribution of $\delta \theta_B$ for OMC-1 shows a clear peak at 0-10$^\circ$, and gradually declines toward 90$^\circ$; for OMC-2, $\delta \theta_B$ has a nearly flat distribution, again indicating that the $B$-fields on the two scales are apparently decoupled; for OMC-3, the distribution of $\delta \theta_B$ depicts that the $B$-field orientation almost do not change across the two scales.

\subsection{Relations between B-field and filamentary structures}
\begin{figure*}[ht]
\centering
\includegraphics[width=1.0\columnwidth]{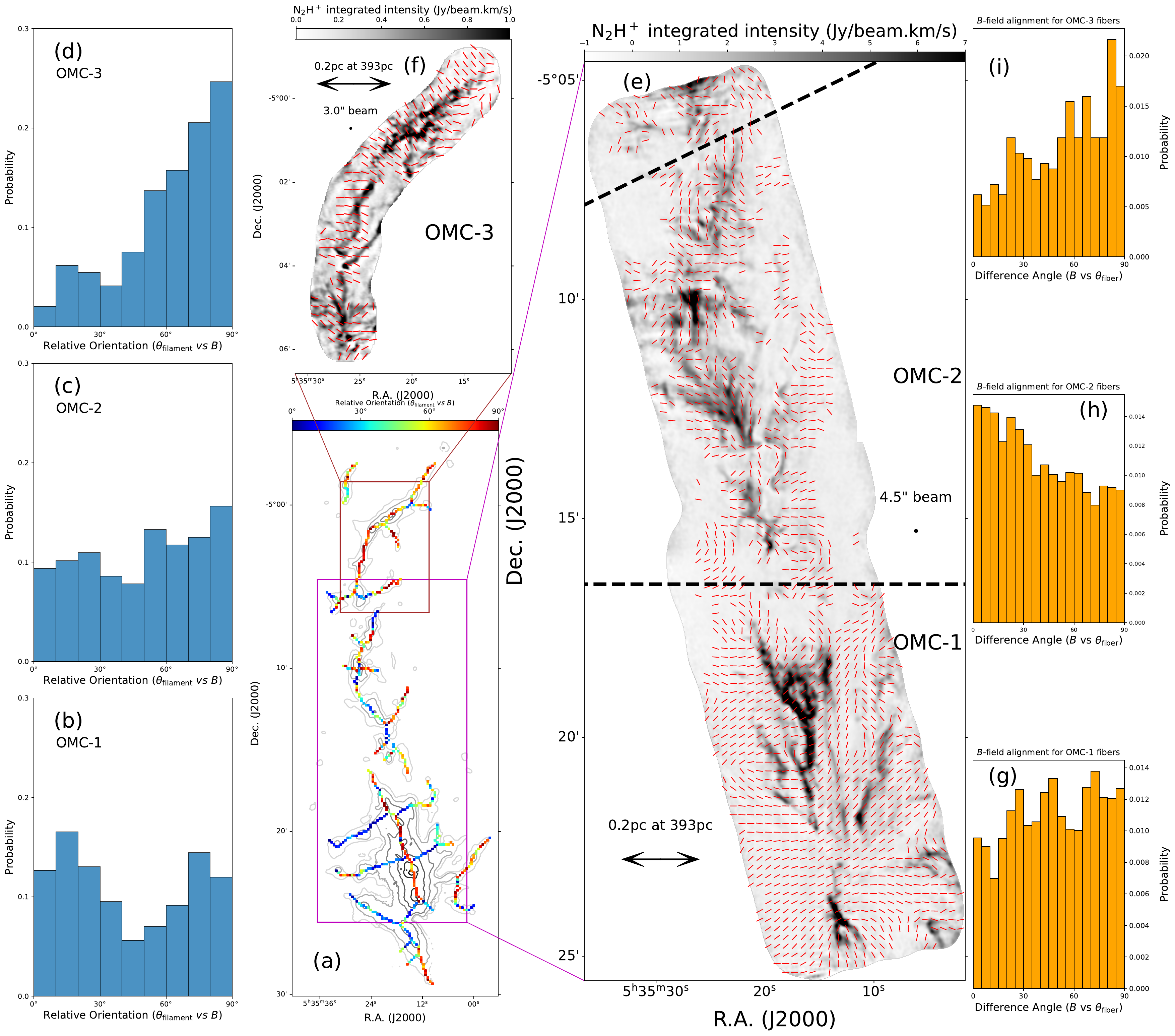}
\caption{Panel (a): Colored points show locations along the skeleton of each filament where the filament orientation is compared with the $B$-field orientation, and the color scale denotes the derived difference angle between the two orientations, as indicated by a color bar on the top; the 850 $\mu$m total intensity is shown in contours with levels at a ${\rm log_{10}}$ scale (${\rm mJy\ beam^{-1}}$), starting from 2.2 and continuing at steps of 0.5. Panels (b), (c), and (d): Histograms of the difference angles between the filament skeleton and $B$-field orientations for OMC-1, OMC-2, and OMC-3. Panel (e): Red vectors plotted at an interval of 12$\arcsec$ show the $B$-field orientations derived by the JCMT POL-2, and the gray scale image shows the $\rm N_2H^+$ (1-0) velocity integrated emission in OMC-1 and OMC-2 \citep{2018A&A...610A..77H}. Panel (f): the same as Panel (e), but for OMC-3, and the $\rm N_2H^+$ data are taken from  \citep{2020MNRAS.497..793Z}. Panels (g), (h), and (i) show the histograms of the difference angles between the $\rm N_2H^+$ fibers and POL-2 $B$-field orientations in OMC-1, OMC-2, and OMC-3, respectively.}
\label{filISF}
\end{figure*}

To investigate how the $B$-field orientation is aligned with the filamentary structures in the ISF, we employed the \texttt{filfinder} algorithm \citep{2015MNRAS.452.3435K} to extract filament skeletons. Figure \ref{filISF}(a) shows the derived skeletons along the main filament, the branches connected to the main filament, and some minor structures detached from the main filament. To quantify the filament orientations, we utilize the Principle Component Analysis (PCA) method on 10 adjacent pixels of the skeletons to determine the PA of the filaments at each position. We then compute the difference angles between the filament and $B$-field orientations. In Figure \ref{filISF}(a), the color scale of the skeletons visualizes the spatial distribution of the difference angles. Figures \ref{filISF}(b-d) show the histograms of the difference angles for OMC-1, OMC-2, and OMC-3, respectively. Along the filamentary cloud, three drastically different distributions for the relative orientation between the $B$-fields and filaments are seen: a bimodal distribution for OMC-1, nearly flat distribution for OMC-2, and a distribution with a predominant single peak at 90$^\circ$ for OMC-3. From the skeleton color scale representing the difference angles (Figure \ref{filISF}a), we can see that the bimodal distribution in OMC-1 is due to a combined effect that along the main filament, the $B$-field orientation is perpendicular to the filament axis, while along the relatively low density branches, the $B$-field orientation is parallel to the branch axis; on the other hand, for the distribution of relative orientation in OMC-3, a tail toward 0$^\circ$ is mostly attributed to OMC-3 South.

Molecular filaments may have complex internal structures, such as intertwined filamentary bundles or fibers. We identified the fibers with the filfinder algorithm from ${\rm N_2H^+}$ maps (see Appendix \ref{app_strc}). Figure \ref{filISF}(e) shows a comparison between the $B$-field orientations derived from our POL-2 observations and the fiber structures revealed by the combined ALMA and IRAM 30m ${\rm N_2H^+}$ (1-0) observations of OMC-1, 2 \citep{2018A&A...610A..77H}. Such a comparison for OMC-3 is shown in Figure \ref{filISF}(f), where the ALMA ${\rm N_2H^+}$ (1-0) data were taken from \citet{2020MNRAS.497..793Z}.  We calculate the difference angles between the fiber and $B$-field orientations, as shown in Figure \ref{filISF}(g-i). In OMC-1, the fibers tend to be perpendicular to the $B$-field; this is not difficult to understand considering the bimodal distribution for the relative orientation between the $B$-field and filaments (Figure \ref{filISF}b) and here the fibers traced by the ${\rm N_2H^+}$ emission represent the high-density part of the filaments. In OMC-3, the fibers are clearly perpendicular to the $B$-field, consistent with the distribution of relative orientation between the $B$-field and filaments. Interestingly, the fibers in OMC-2 appear to be predominantly parallel to the $B$-field, in contrast to the random distribution of relative orientation between the $B$-field and filaments.

\section{Discussion and Summary} \label{sec:disc}
\subsection{A tale of three: gravitational, turbulent, and magnetic interpretations for OMC-1, 2, and 3, respectively}
We have presented JCMT POL-2 dust polarization observations of a remarkable molecular filament containing OMC-1, OMC-2, and OMC-3 in the Orion ISF. Combing the POL-2 data with the \emph{Planck} and CARMA polarization observations, we clearly see how the $B$-fields vary from the large ($\sim$0.6 pc) to intermediate ($\sim$0.03 pc) and small ($\sim$0.005) scales: for OMC-1, the $B$-field retains its mean orientation on all the scales, with some local variations on intermediate to small scales; for OMC-2, the $B$-fields on different scales are apparently decoupled, showing relatively disordered morphology on the intermediate and small scales; for OMC-3, the $B$-field shows a uniform morphology and the orientation does not change all the way from the large to intermediate and small scales. A natural and straightforward interpretation of such $B$-field morphologies, in particular their variation across different scales, is that the $B$-field in OMC-1 is channeling the gas accretion from the ambient medium to the filament, but as the mass continues to grow, forming massive dense cores within the filament, gravity overcomes the magnetic force, pulling the $B$-field into an hour-glass shape \citep[see also][]{2017ApJ...842...66W,2017ApJ...846..122P}. The $B$-field in OMC-2 appears highly twisted on the intermediate and small scales, suggesting that turbulence is dominating over the $B$-field; the $B$-field in OMC-3, especially OMC-3 North, has a nearly uniform morphology from large to small scales, indicating that the $B$-field is strong enough to dominate the gas dynamics \citep[e.g.,][]{2001ApJ...546..980O}. Below we test this simple interpretation by comparing the orientations between the $B$-fields and the dense gas structures.
    
Filamentary clouds naturally define an axis to be compared to the $B$-field, and such a comparison for ISF reveals again a trio: bimodal for OMC-1, random for OMC-2, and perpendicular for OMC-3 (Figures \ref{filISF}b-d). It immediately renders strong support to the above ternary interpretation. The bimodal distribution for the relative orientation between the $B$-field and filaments in OMC-1 is clearly correlated to the gas density, with the high-density filament skeleton perpendicular to the $B$-field and low-density skeletons parallel to the $B$-field, consistent with the scenario that the $B$-field is channeling gas flows toward the high-density filament \citep[e.g.,][]{2017ApJ...842...66W,2020NatAs...4.1195P,2021MNRAS.507.5641G}. Such a correlation is strengthened by looking into the filament internal structures, i.e., the ${\rm N_2H^+}$ fibers: as the high-density part of the filament, the fibers are preferentially perpendicular to the $B$-field (Figure \ref{filISF}g). In OMC-2, the random distribution is apparently a consequence of the disordered nature of the $B$-field structure. Very interestingly, the fibers in OMC-2 are largely parallel to the $B$-field (Figure \ref{filISF}h), showing a pattern that is consistent with the results of simulations of super-Alfvenic turbulence \citep[see, e.g., Figure 2 and Figure 3 in][]{2001ApJ...559.1005P}, suggesting that turbulence is dynamically more important than the $B$-field in OMC-2. For OMC-3, the $B$-field is simply perpendicular to both the filament (Figure \ref{filISF}d) and fibers (Figure \ref{filISF}i), indicating that the $B$-field is strong enough to counteract gravity and turbulence.

\subsection{The B-field strength estimates}
To further quantify the impact of the $B$-field on the dynamical evolution of the filament, it is desirable to estimate the $B$-field strength. However, deriving the $B$-field strength with the David-Chandrasekhar-Fermi (DCF) method \citep{1951PhRv...81..890D,1953ApJ...118..113C} or its variants is subject to large uncertainty and in some cases is not applicable \citep{2021ApJ...919...79L,2022ApJ...925...30L,2022FrASS...9.3556L,2022MNRAS.514.1575C}. First, the method requires calculation of the polarization angle dispersion due to turbulent disturbance, or decomposing the $B$-field into turbulent and ordered components and calculating their ratio. This step is not always feasible, especially when the $B$-field structure is complicated. Second, under the assumption of energy equipartition between turbulence and the turbulent $B$-field, and adopting a gas density and turbulent velocity dispersion obtained from other observations, the plane-of-sky (PoS) $B$-field strength can be derived. It should be noted that the energy equipartition assumption may not be valid when the $B$-field is weak. The estimates of the gas density and turbulent velocity dispersion often suffer large uncertainties. Nevertheless, the method has been widely used. Several such estimates for the sources in the ISF exist in the literature, and the results vary a lot, ranging from 0.3 to 6.6 mG for OMC-1 and from 0.13 to 0.64 mG for OMC-3 \citep{2005ApJ...626..959M,2007AJ....133.1012V,2009ApJ...696..567H,2009ApJ...706.1504H,2013ApJ...777..112P,2017ApJ...846..122P,2019ApJ...872..187C,2021ApJ...908...98G,2021ApJ...913...85H,2022MNRAS.514.3024L,2022A&A...659A..22Z}. Here we try with the best effort to estimate the PoS $B$-field strengths in the three regions with the new data, obtaining 0.45, 0.25, and 0.37 mG for OMC-1, 2, and 3, respectively (see Appendix \ref{app_Bstrength}). 

Given the aforementioned cautions and uncertainties, we only make a comparative rather than more detailed quantitative analysis based on the derived $B$-field strengths. It is worth mentioning that OMC-1 has a width about two times greater than OMC-2 and 3 under a same column density threshold, resulting in a volume density in this region slightly lower than that in the latter two. But about half of the gas mass in OMC-1 is attributed to the central high-density part with a width of $\sim0.06$ pc, and within that area the average volume density reaches $2.5\times10^6$ cm$^{-3}$. Therefore, though the $B$-field in OMC-1 is stronger, considering a much greater mass and central density, it is completely plausible that gravity is overwhelmingly more important in this region. OMC-2 and 3 have almost the same mass and non-thermal velocity dispersion (see Appendix \ref{app_Bstrength}), while the $B$-field in OMC-3 is stronger than that in OMC-2; the relative $B$-field strength of the two regions is at least compatible with the interpretation that turbulence in OMC-2 and $B$-field in OMC-3 is taking a leading role.

\subsection{The impact of magnetic field on star formation}
\begin{figure}[htbp]
\centering
\includegraphics[width=0.8\columnwidth]{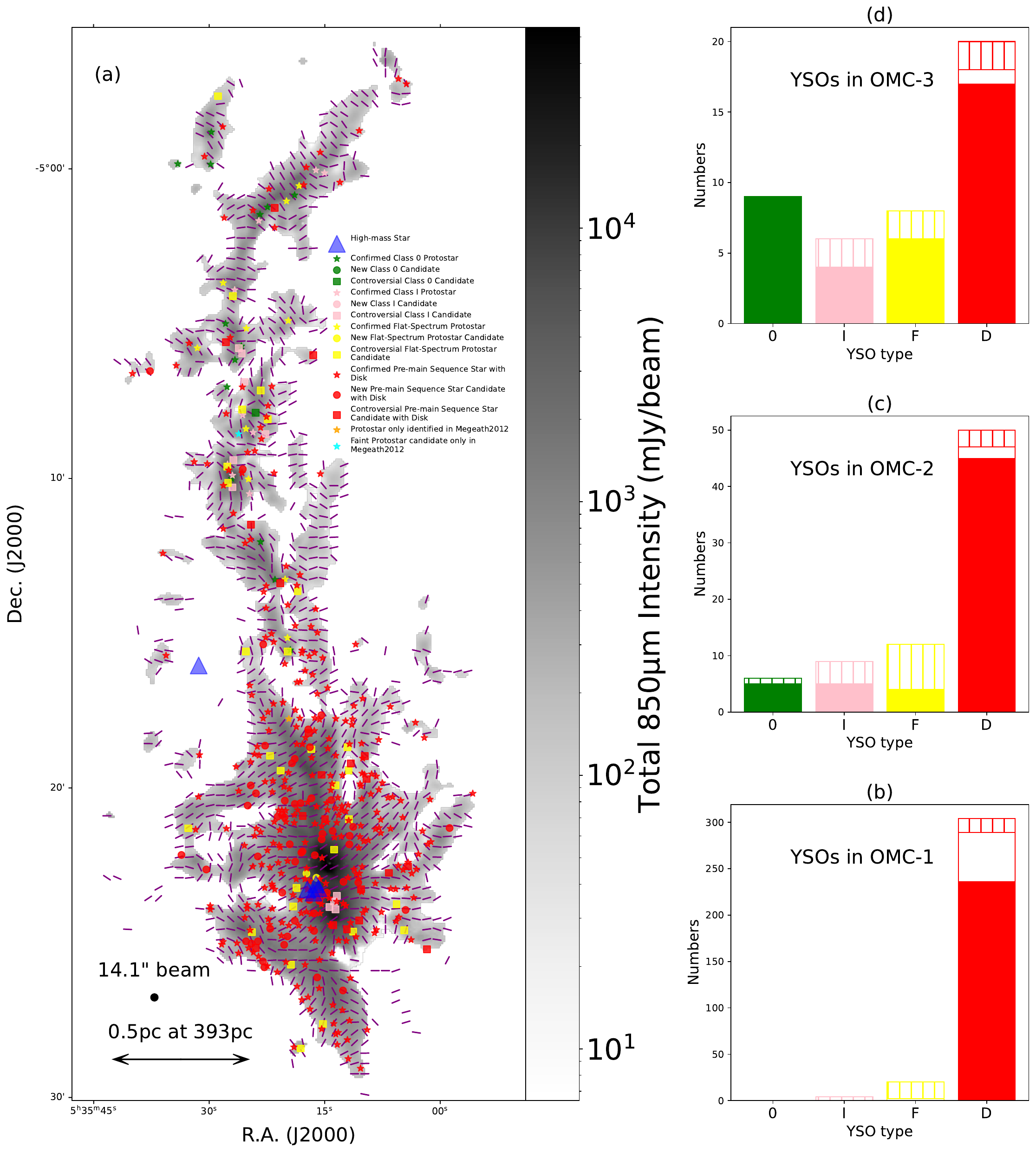}
\caption{Panel (a): Distribution of detected YSOs overlaid on the $B$-field orientation maps. All the YSO candidates and high-mass stars are taken from the literature \citep{2012AJ....144..192M, 2016ApJS..224....5F, 2019A&A...622A.149G}. Purple vectors at a 20$\arcsec$ interval indicate the $B$-field orientations observed by the JCMT POL-2. The Class 0, Class I, flat-spectrum sources, and pre-main sequence stars with disks are denoted in green, pink, yellow, and red colors, respectively. For each YSO type, the confirmed ones that are consistent in different literature are represented by star symbols; newly discovered candidates by \citet{2019A&A...622A.149G} are indicated with filled circles; controversial candidates, showing inconsistencies in different literature, are marked with filled squares. Panels (b), (c), and (d): Histograms of the four YSO types in OMC-1, OMC-2, and OMC-3, respectively; Class 0, Class I, flat-spectrum sources, and pre-main sequence stars with disks are labeled as `0', `I', `F', and `D', respectively, with the same colors as in Panel (a); filled histograms represent confirmed YSOs, hollow histograms depict newly discovered YSOs, and vertical gridded histograms indicate controversial YSOs.}
\label{fig:Protostar}
\end{figure}

Given the markedly different $B$-field properties across the three regions in the ISF, it is of great interest to examine how the star formation activity is affected. We collect a catalog of Young Stellar Objects (YSOs), which are classified into Class 0, Class I, flat-spectrum, and disk-bearing pre-main-sequence stars, based on the works of \citep{2012AJ....144..192M,2016ApJS..224....5F,2019A&A...622A.149G}. Figure \ref{fig:Protostar} shows all the YSOs in the OMC-1, 2, and 3, and the statistics of each type in each of the three regions. The star formation activity in OMC-1 is far more vigorous and complicated than that in OMC-2 and 3. OMC-1 is the only region of the three forming high-mass stars, containing several well-known high-mass protostellar objects. It is located behind the luminous Trapezium cluster, which is the central part of the Orion Nebular Cluster (ONC). The collected YSOs in this region is completely dominated by the disk sources, and are heavily contaminated by the foreground ONC sources \citep{2000AJ....120.3162L,2021ApJ...923..221O}. Here we focus on the comparison between OMC-2 and OMC-3. From Figure \ref{fig:Protostar}, OMC-2 has a higher fraction of disk sources (50/78) than OMC-3 (20/43), indicating a younger age of the cluster in OMC-3. The total number of YSOs in OMC-2 is higher than that in OMC-3. Note that the mass, mean density, and non-thermal velocity dispersion in the two regions are almost the same, and the only appreciable difference lies in the $B$-field geometries and the relative orientation between the $B$-fields and filaments/fibers. Therefore the differing YSO populations in the two regions are mostly likely due to the $B$-field effect, providing compelling evidence that a dynamically more important $B$-field leads to slower (or delayed) and less efficient star formation in molecular clouds.

To summarize, concerning which mechanism is shaping the dynamics of molecular clouds on $\sim$0.01 to 1 pc scales, each of the three clouds (OMC-1, 2, and 3 in the Orion ISF) seems to be telling a different story based on our JCMT POL-2 observations along with the \emph{Planck} and CARMA data. Therefore it is probably an over-simplified interpretation to claim that either magnetic field or turbulence is universally more important in molecular cloud evolution and star formation. By comparing the YSO populations in OMC-2 and 3, we find evidence that a strong $B$-field could make star formation relatively slower and less efficient. \citet{2019ApJ...871...98Z} carried out MHD simulations of sub-Alfv\'{e}nic molecular clouds, focusing on the $B$-field orientation variation across various scales. They found that on small ($<$0.1 pc) scales, the cores are super-Alfv\'{e}nic, as a consequence of turbulent energy concentration induced by gravity, and thus the $B$-field on small scales exhibits a wide range of deviation in orientation from that on large scales. If one takes an average $B$-field orientation for each of the dense cores in the CARMA maps (Figure \ref{figveccom}c-e) and compare to the $B$-field revealed by \emph{Planck}, the offset distribution could be to some extent consistent with the work of \citet{2019ApJ...871...98Z}. However, a detailed comparison shows that the cross-scale correlation in $B$-field orientation (Figure \ref{figveccom}f, g) is distinctly different from region to region, certainly not random in OMC-1 and 3. The observed relation between the $B$-field and filament/fiber orientations and the star formation activity variation further suggest a tale-of-three interpretation of the three regions regarding the interplay between gravity, $B$-field, and turbulence.

\begin{acknowledgments}
This work is supported by the National Natural Science Foundation of China (NSFC) grants No. 12425304 and No. U1731237, the National Key R\&D Program of China with No. 2023YFA1608204 and No. 2022YFA1603103, and the science research grant from the China Manned Space Project with no. CMS-CSST-2021-B06. J.W. thanks Chao Zhang, Nan-Nan Yue, Di Li, Qizhou Zhang for generously providing us the ${\rm N_2H^+}$ data that covered OMC-2 and OMC-3 in Orion for the analysis of fiber structures. C.E. acknowledges the financial support from the grant RJF/2020/000071 as a part of the Ramanujan Fellowship awarded by the Science and Engineering Research Board (SERB). The James Clerk Maxwell Telescope is operated by the East Asian Observatory on behalf of The National Astronomical Observatory of Japan; Academia Sinica Institute of Astronomy and Astrophysics; the Korea Astronomy and Space Science Institute; Center for Astronomical Mega-Science (as well as the National Key R\&D Program of China with No. 2017YFA0402700). Additional funding support is provided by the Science and Technology Facilities Council of the United Kingdom and participating universities in the United Kingdom, Canada, and Ireland. Additional funds for the construction of SCUBA-2 and POL-2 were provided by the Canada Foundation for Innovation.
\end{acknowledgments}

\appendix

\section{Identified fiber Structures}\label{app_strc}
\begin{figure}[htbp]
\centering
\includegraphics[width=1.0\columnwidth]{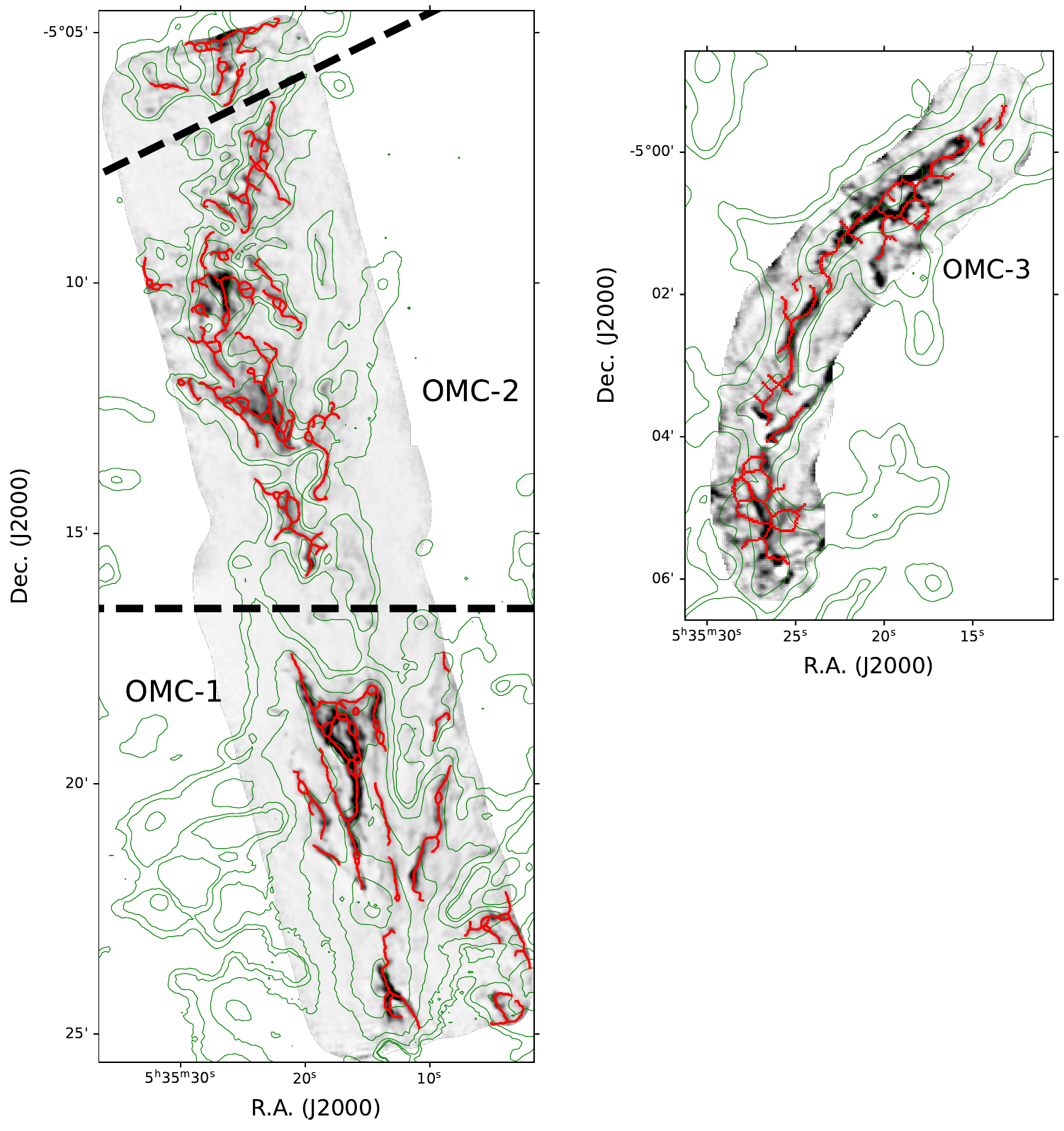}
\caption{Gray scale images show the velocity integrated $\rm N_2H^+$ (1-0) emissions, overlaid with the extracted fibers shown in red lines and the total 850 $\mu$m intensity shown in green contours. The left panel shows the OMC-1, 2 region, with the $\rm N_2H^+$ data taken from \citep{2018A&A...610A..77H}, and the right panel shows OMC-3, with the $\rm N_2H^+$ data taken from \citep{2020MNRAS.497..793Z}.}
\label{fig:Fiber_ISF}
\end{figure}

The fibers within the filament are extracted using the filfinder algorithm from the $\rm N_2H^+$ (1-0) velocity integrated emission maps. Figure \ref{fig:Fiber_ISF} shows a comparison between the derived fibers, the $\rm N_2H^+$ emission, and the total 850 $\mu$m emission.

\section{Details for B-field strength calculation}\label{app_Bstrength}
In the DCF assumption, the PoS $B$-field strength of the molecular cloud is estimated by interpreting the observed deviation of polarization angles from a mean polarization angle distribution as a result of Alfv\'{e}n waves induced by turbulent perturbations. i.e.:
\begin{equation}
B_0={\sigma_v}\sqrt{\mu_0\rho}(\frac{\delta B}{B_0})^{-1}
\label{eq:eq_B}
\end{equation}
where $\sigma_v$ represents the turbulence-induced velocity dispersion which could approximately equate to the non-thermal velocity dispersion, $\rho$ denotes the gas mass density. ${\delta B}/{B_0}$ denotes the turbulent-to-ordered magnetic field ratio.

To obtain the mass density, we modeled the three star-forming clouds within the Orion A ISF as cylindrical filaments. We use the column density map at $\sim$ 8$\arcsec$ resolution produced by \citet{2021A&A...651A..36S} to measure the mass of the three regions, obtaining $\sim$660, $\sim$250, and $\sim$260 $M_{\sun}$ for OMC-1, 2, 3. The dimensions of the three regions are measured to be approximately 0.93 pc $\times$ 0.23 pc for OMC-1, 1.0 pc $\times$ 0.1 pc for OMC-2, and 1.0 pc $\times$ 0.1 pc for OMC-3. Assuming a cylinder geometry lying in the plane of sky, the volume densities are found to be $\sim{\rm 2.4\times10^{5}}$, $\sim{\rm 4.5\times10^{5}}$, and $\sim{\rm 4.7\times10^{5}\ cm^{-3}}$ for OMC-1, 2, and 3, respectively.

To estimate the velocity dispersion in the ISF, we utilized the $\rm NH_3$ (1,1) observation data from the Green Bank Ammonia Survey \citep[GAS,][]{2017ApJ...843...63F} with a resolution of 36$\arcsec$. To extract the non-thermal velocity dispersion, we subtracted the thermal components of the observed velocity dispersion with the temperature map provided by \citet{2021A&A...651A..36S}. Our analysis revealed that the mean non-thermal velocity dispersion in OMC-1, OMC-2, and OMC-3 is 0.90 $\rm km\ s^{-1}$, 0.38 $\rm km\ s^{-1}$, and 0.41 $\rm km\ s^{-1}$, respectively.

The turbulent-to-ordered magnetic field ratio ${\delta B}/{B_0}$ is determined by the dispersion of polarization angles. However, quantifying the turbulent $B$-field components could have bias due to the effects of non-turbulent field structure in dense clouds. So the angular dispersion function method has been developed to reduce the bias. Moreover, by considering the effect of signal integration along the line of sight and within the beam in the analysis, \citet{2009ApJ...706.1504H} proposed the Autocorrelation Function (ACF) form to precisely derive the turbulent-to-ordered magnetic field ratios. The angular dispersion function could be expressed as:

\begin{equation}
1 - \langle \cos \lbrack \Delta \Phi (l)\rbrack \rangle \simeq \frac{1}{N} \frac{\langle {\delta}B^2 \rangle}{\langle B_0^2 \rangle} \times \lbrack 1 - e^{-l^2/2(\delta^2+2W^2)}\rbrack + a_2 l^2
\label{eql}
\end{equation}
where $N$ is the number of turbulent cells probed by the telescope beam. $\Delta \Phi (l)$ represents the position angle differences of two vectors at a distance $l$, $a_2$ signifies the slope of the second-order term in the Taylor expansion.
\begin{equation}
N = \frac{(\delta^2 + 2W^2)\Delta^\prime}{\sqrt{2\pi}\delta^3},
\end{equation}
$W$ denotes the beam radius (6.0$\arcsec$ for JCMT 850$\mu$m observations), $\Delta^\prime$ depicts the cloud depth and $\delta$ stands for the turbulent correlation length. 

\begin{figure}[htbp]
\centering
\includegraphics[width=1.0\columnwidth]{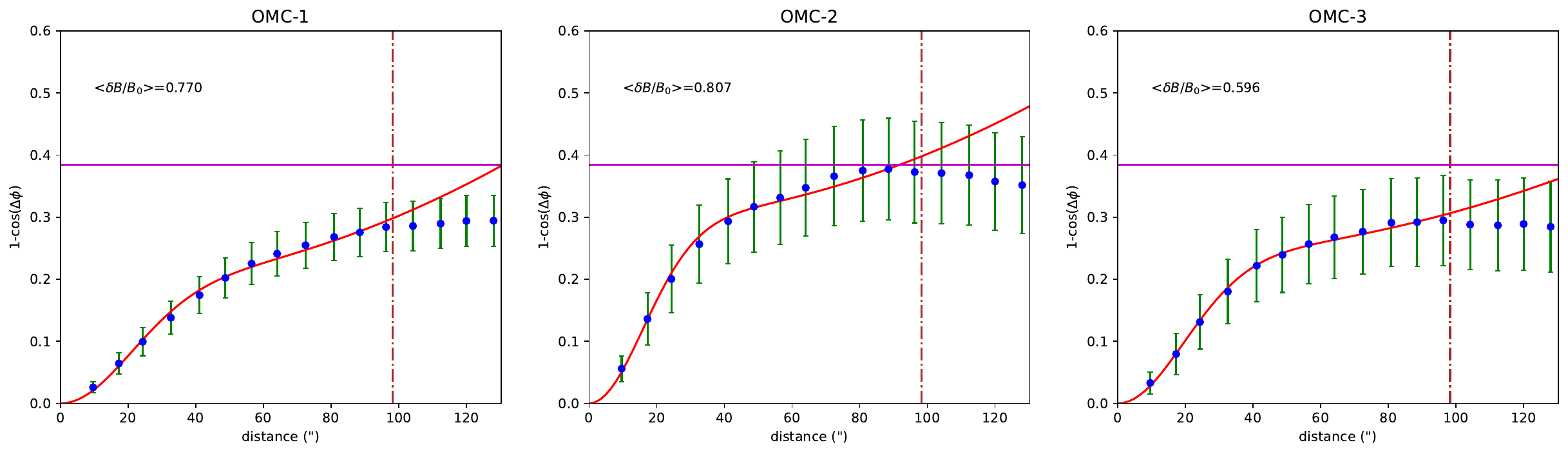}
\caption{Fitting results of the ACF for OMC-1, 2, and 3 from left to right. For each panel, blue filled circles with error bars denote data points derived from the polarization observations; the best fitting result is shown by a red solid line; a horizontal magenta line marks the value expected for a random field \citep[52°,][]{2010ApJ...716..893P}; a vertical brown dashed line marks the right boundary of the points to be fitted.}
\label{fig:ACF_OMC123}
\end{figure}

Setting cloud depths to 0.23pc, 0.1pc and 0.1pc for OMC-1, OMC-2, and OMC-3 respectively, we derived the ACF of the three clouds with the JCMT POL-2 polarization vecotors (2952 vectors in OMC-1, 1118 vectors in OMC-2, 890 vectors in OMC-3). Equation \ref{eql} is valid when $l$ is not too big compared to a few times of $W$ \citep{2009ApJ...706.1504H}. In addition, we have a polarization map with a finite size, and thus the number of polarization detections on which the ACF could be derived at high intensities decreases as $l$ increases, leading to degrading statistics for the data points on large $l$. We therefore limit our fitting to the data points with $l<100\arcsec$. In Figure \ref{fig:ACF_OMC123}, the fitting results revealed that OMC-3 has the smallest $\delta B/B_0$ value of 0.596 and OMC-2 has the largest $\delta B/B_0$ value of 0.807. While OMC-1 has a $\delta B/B_0$ value of 0.770. We also obtain $\delta$=4.39, 3.15, and 4.08 mpc for OMC-1, 2, and 3, respectively, with the fitting. We note that $\delta$ cannot be resolved with a telescope beam of 27 mpc ($14''$ at a distance of 393 pc). Such an issue occurs in other works of applying the ACF fitting to dust polarization data \citep[e.g.,][]{2009ApJ...706.1504H,2013ApJ...779..182Q}. Thus the inferred $\delta$ is more like a numerical artifact from the fitting, and the turbulence correlation scale is still to be explored. We finally estimated the strength of the PoS component of the $B$-field for OMC-1, OMC-2, and OMC-3 as 0.45\,mG, 0.25\,mG, and 0.37\,mG, respectively.

\bibliography{OrionISF_AAS_JintaiWu}{}
\bibliographystyle{aasjournal}
\end{document}